\documentstyle[sprocl]{article}

\def\be{\begin{equation}}
\def\ee{\end{equation}}
\def\bea{\begin{eqnarray}}
\def\eea{\end{eqnarray}}

\begin{document}

\title{ACOUSTICS IN BOSE--EINSTEIN CONDENSATES AS AN EXAMPLE OF
 BROKEN LORENTZ SYMMETRY}

\author{MATT VISSER, CARLOS BARCEL\'O}

\address{Washington University, Saint Louis, MO 63130-4899, USA\\ 
E-mail: visser@kiwi.wustl.edu; carlos@hbar.wustl.edu} 

\author{STEFANO LIBERATI}

\address{University of Maryland, College Park, MD 20742--4111, USA\\
E-mail: liberati@physics.umd.edu}


\maketitle\abstracts{
To help focus ideas regarding possible routes to the breakdown of
Lorentz invariance, it is extremely useful to explore concrete
physical models that exhibit similar phenomena. In particular,
acoustics in Bose--Einstein condensates has the interesting property
that at low-momentum the phonon dispersion relation can be written in
a ``relativistic'' form exhibiting an approximate ``Lorentz
invariance''. Indeed all of low-momentum phonon physics in this system
can be reformulated in terms of relativistic curved-space quantum
field theory. In contrast, high-momentum phonon physics probes regions
where the dispersion relation departs from the relativistic form and
thus violates Lorentz invariance. This model provides a road-map of at
least one route to broken Lorentz invariance. Since the underlying
theory is manifestly physical this type of breaking automatically
avoids unphysical features such as causality violations. This model
hints at the type of dispersion relation that might be expected at
ultra-high energies, close to the Planck scale, where quantum gravity
effects are suspected to possibly break ordinary Lorentz invariance.
\\[5pt]
{\emph{Presented at CPT01; the Second Meeting on CPT and Lorentz
Symmetry; \\ Bloomington, Indiana; 15--18 Aug 2001.}}
\\ 
hep-th/0109033; 5 September 2001;  \LaTeX-ed \today.}

\def\d{{\mathrm{d}}}
\def\Barcelo{Barcel\'o}

\section{Introduction}
\label{s:introduction}

Physics is often studied by analogy, and the last few years has seen
the development of a number of deep and fruitful analogies between
various condensed matter systems and relativistic physics.  In this
article we will focus on acoustic propagation in Bose--Einstein
condensates,~\cite{barcelo,garay} which has the interesting property
that the low-energy spectrum exhibits an effective Lorentz invariance
(in terms of the speed of sound) while at high energies the dispersion
relation turns over and becomes Newtonian.~\cite{barcelo}

During this conference on CPT invariance and Lorentz symmetry there
have been numerous reports, based on as yet partially developed
theories of quantum gravity, that hint at a breakdown of Lorentz
invariance at or near the Planck scale. On the other hand, it is
standard lore that Lorentz symmetry breaking leads to severe causality
problems. It is therefore useful to have at least one concrete
physical model of Lorentz symmetry breaking at hand which is
guaranteed to be physically consistent. Even if the symmetry breaking
believed to arise from quantum gravity does not follow this particular
pattern, exhibiting such a model unambiguously settles an important
matter of principle. Additionally, the type of dispersion relation
arising in this model (and many other related analog
models~\cite{didactic,viscosity}) gives an important hint as to what
type of modified dispersion relation to look for at Planck energies.

\section{From BEC to Lorentzian geometry}
\label{s:bec2lorentz}

Bose--Einstein condensates are described by the nonlinear
Schrodinger equation (Gross--Pitaevskii equation):
\be
- i \hbar \; \partial_t \psi(t,\vec x) = - {\hbar^2\over2m}\nabla^2
\psi(t,\vec x) + \lambda \; ||\psi||^2 \; \psi(t,\vec x).
\ee
(We have suppressed the externally applied trapping potential for
algebraic simplicity. For technical details see {\Barcelo}
{\emph{et~al.}}~\cite{barcelo} That reference also contains an
extensive background bibliography.) Now use the Madelung
representation to put the Schrodinger equation in ``hydrodynamic''
form:
\be
\psi = \sqrt{\rho} \; \exp(-i\theta\; m/\hbar).
\ee
Take real and imaginary parts: The imaginary part is a continuity
equation for an irrotational flow of velocity $\vec v \equiv
\nabla\theta$ and the real part is a Hamilton--Jacobi equation (its
gradient leads to the Euler equation). Specifically:
\be
\partial_t \rho + \nabla\cdot(\rho \; \nabla \theta) = 0.
\ee
\be
\frac{\partial}{\partial t} \theta + {1\over2}(\nabla \theta)^2 
+ {\lambda \; \rho \over m}
- {\hbar^2\over2m^2}\; 
{\Delta\sqrt\rho\over\sqrt\rho}= 0.
\ee
That is, the nonlinear Schrodinger equation is completely equivalent
to irrotational inviscid hydrodynamics with an enthalpy
\be
h = \int {\d p \over \rho} = {\lambda \; \rho\over m}, 
\ee
plus a peculiar self-interaction:
\be
V_Q = - {\hbar^2\over2m^2}\; {\Delta\sqrt\rho\over\sqrt\rho}= 0.
\ee
The equation of state for this Madelung fluid is
$p = {\lambda \;\rho^2\over2m}$; so that $ {\d p\over\d\rho} 
= {\lambda \; \rho\over m}$.
To now extract a Lorentzian geometry, linearize around some
background. In the low-momentum limit it is safe to neglect $V_Q$. It
is a by now standard result that the {\emph{phonon} is described by a
massless minimally-coupled scalar that satisfies the d'Alembertian
equation in the effective (inverse) background-dependent
metric~\cite{viscosity,unruh}
\be
g^{\mu\nu}(t,\vec x) \equiv 
{\rho_0\over c_s}
\left[ \matrix{-1&\vdots&-v_0\cr
               \cdots\cdots&\cdot&\cdots\cdots\cdots\cdots\cr
	       -v_0&\vdots&(c_s^2 \;{\mathbf{I}} - v_0 \otimes v_0)\cr } 
\right].	       
\ee
Here
\be
c_s^2 \equiv {\lambda \; \;\rho_0\over m}; 
\qquad\qquad
v_0 = {\nabla \theta_0}.
\ee
It cannot be overemphasized that low-momentum phonon physics is
completely equivalent to quantum field theory in curved spacetime.
That is, everything that theorists have learned about curved space QFT
can be carried over to this acoustic system, and conversely acoustic
experiments can in principle be used to experimentally investigate
curved space QFT.  In particular, it is expected that acoustic black
holes (dumb holes) will form when the condensate flow goes supersonic,
and that they will emit a thermal bath of (acoustic) Hawking radiation
at a temperature related to the physical acceleration of the
condensate as it crosses the acoustic
horizon.\cite{barcelo,garay,viscosity,unruh} 

For completeness we mention that the metric is
\be
g_{\mu\nu}(t,\vec x) 
\equiv \; {\rho_0\over c_s} \;
\left[ \matrix{-(c_s^2-v_0^2)&\vdots&-{\vec v}_0\cr
               \cdots\cdots\cdots\cdots&\cdot&\cdots\cdots\cr
	       -{\vec v}_0&\vdots& \mathbf{I}\cr } \right]
\ee
so the space-time interval can be written
\be
\d s^2 =
{\rho_0\over c_s} \;
\left[
- c_s^2 \; \d t^2 + ||\d\vec x - \vec v_0 \; \d t||^2 
\right].
\ee

\section{Characteristics}
\label{s:characteristics}

But there is a bit of a puzzle here: We started with the nonlinear
Schrodinger equation. That equation is parabolic, so we know that the
characteristics move at infinite speed.  How did we get a hyperbolic
d'Alembertian equation with a finite propagation speed?  The subtlety
resides in neglecting the higher-derivative term $V_Q$. To see this,
keep $V_Q$, and go to the eikonal approximation. One obtains the
dispersion relation~\cite{barcelo}
\be
\left(\omega - \vec v_0 \cdot \vec k\right)^2 =
{c_s^2 \;k^2} +\left({\hbar\over2m} k^2\right)^2.
\ee
This is the curved-space generalization of the well-known Bogolubov
dispersion relation. Equivalently
\be
\omega=  \vec v_0 \cdot \vec k  +
\sqrt{ {c_s^2 \;k^2} +\left({\hbar\over2m} k^2\right)^2 }.
\ee
The group velocity is
\be
{\vec v}_g = {\partial\omega\over\partial\vec k} =  
\vec v_0  
+
{ \left(c_s^2+{\hbar^2 \over 2 m^2}k^2\right) 
\over 
\sqrt{c_s^2 k^2+\left({\hbar \over 2 m}\;k^2\right)^2} }
\; \vec k,
\ee
while for the phase velocity
\be
\vec v_p = {\omega\; \hat k\over||k||} =  (v_0 \cdot \hat k) \; \hat k  
+
\sqrt{c_s^2+{\hbar^2 \;k^2\over 4 m^2}} \;
\; \hat k.
\ee
Both group and phase velocities have the appropriate relativistic
limit at {\emph{low}} momentum, but then grow without bound at
{\emph{high}} momentum, leading to an infinite signal velocity and the
recovery of the parabolic nature of the differential equation at high
momentum. ($k \gg k_c \equiv m\; c_s/\hbar$; equivalently in terms of
the acoustic Compton wavelength $\lambda \ll \lambda_C \equiv
\hbar/(m\; c_s)$.)

It is amusing, and perhaps somewhat surprising, to notice that while
at low momentum the Bogolubov dispersion relation is (approximately)
relativistic, at large momenta ($k \gg k_c$) the dispersion relation
again takes on Newtonian form, complete with ``rest mass''
contribution
\be
\omega(k) = 
{\hbar \;k^2\over2m} + \vec v_0 \cdot \vec k  + {m \; c_s^2\over\hbar} 
+ O(k^{-2}).
\ee
%

\section{Discussion}
\label{s:discussion}

The key message to take from this article is that BECs are examples of
physically realizable systems with ``broken'' Lorentz invariance. The
way to preserve causality in this system is clear --- there is an
intrinsic preferred frame in this model and one should quantize by
applying equal time commutators in the inertial frame in which the
original nonlinear Schrodinger equation is written down. In
particular, field commutators need not generically vanish identically
outside the lightcone. In the low-momentum limit, where the physics
decouples from this preferred frame, the commutators are approximately
zero outside the lightcone; this is good enough for an approximate
Lorentz symmetry to arise. (In the preferred frame, at equal times
commutators are exactly zero at disjoint points.)  This is not the
only way of dealing with causality in theories with broken Lorentz
invariance, but it is at least one route that is guaranteed (by its
very construction) to be internally consistent.

It is also useful to remember that BECs are not the only interesting
systems exhibiting broken Lorentz symmetry, in fact condensed matter
physics is littered with ``analog models'' for low-energy Lorentz
invariance.~\cite{didactic} For instance we mention acoustics in the
presence of viscosity,\cite{viscosity} and Lattice phonons.  Finding
an approximate Lorentzian geometry is really just a matter of
isolating a particular degree of freedom that is approximately
decoupled from the rest of the physics and doing a low-momentum
field-theory normal-modes analysis.\cite{normal-modes} In all of the
condensed matter systems we are aware of, the modifications to the
high-energy dispersion relation show up as {\emph{even}} powers of
energy and/or momentum. Ultimately this is due to the fact that the
condensed matter systems considered so far explicitly conserve parity
and time reversal invariance. In contrast some of the models based on
quantum gravity hint at cubic deviations from a quadratic dispersion
relation --- this can often be traced back to some underlying
assumption of an intrinsic violation of parity or time reversal
invariance.

In summary: BECs in particular, and condensed matter systems in
general, provide useful explicit models of an approximate low-energy
Lorentz invariance that is broken by higher-energy ``fundamental''
physics. They provide useful templates for comparison with (and
contrast to) the models of Lorentz symmetry breaking currently
emerging from various quantum gravity scenarios.


\section*{References}

\end{document}